\documentclass[twocolumn]{aastex631}

\usepackage{CJK}

\graphicspath{{./}{figures/}}

\received{XXX}
\revised{YYY}
\accepted{ZZZ}

\usepackage{graphicx}	
\usepackage{amsmath}	
\usepackage{amssymb}	
\usepackage[inline]{enumitem}
\usepackage{gensymb}
\usepackage{array}
\newcolumntype{P}[1]{>{\centering\arraybackslash}p{#1}}
\newcolumntype{M}[1]{>{\centering\arraybackslash}m{#1}}

\usepackage{color}

\shorttitle{Transit Probability, Biases, and Nodal Precession}
\shortauthors{Stephan \& Gaudi 2023}

\begin{document}
\begin{CJK*}{UTF8}{gbsn}

\title{Exoplanet Nodal Precession Induced by Rapidly Rotating Stars:\\Impacts on Transit Probabilities and Biases}

\correspondingauthor{Alexander P. Stephan}
\email{stephan.98@osu.edu}

\author[0000-0001-8220-0548]{Alexander P. Stephan}
\affiliation{Department of Astronomy, The Ohio State University, Columbus, OH 43210, USA}
\affiliation{Center for Cosmology and AstroParticle Physics, The Ohio State University, Columbus, OH 43210, USA}

\author[0000-0003-0395-9869]{B. Scott Gaudi}
\affiliation{Department of Astronomy, The Ohio State University, Columbus, OH 43210, USA}




\begin{abstract}
For the majority of short period exoplanets transiting massive stars with radiative envelopes, the spin angular momentum of the host star is greater than the planetary orbital angular momentum.  In this case, the orbits of the planets will undergo nodal precession, which can significantly impact the probability that the planets transit their parent star. In particular, for some combinations of the spin-orbit angle $\psi$ and the inclination of the stellar spin $i_*$, all such planets will eventually transit at some point over the duration of their precession period. Thus, as the time over which the sky has been monitored for transiting planets increases, the frequency of planets with detectable transits will increase, potentially leading to biased estimates of exoplanet occurrence rates, especially orbiting more massive stars. Furthermore, due to the dependence of the precession period on orbital parameters such as spin-orbit misalignment, the observed distributions of such parameters may also be biased. We derive the transit probability of a given exoplanet in the presence of nodal precession induced by a rapidly spinning host star. We find that the effect of nodal precession has already started to become relevant for some short-period planets, i.e., Hot Jupiters, orbiting massive stars, by increasing transit probabilities by of order a few percent for such systems within the original $Kepler$ field. We additionally derive simple expressions to describe the time evolution of the impact parameter $b$ for applicable systems, which should aid in future investigations of exoplanet nodal precession and spin-orbit alignment.
\end{abstract}


\section{Introduction}\label{sec:intro}

Over recent decades thousands of exoplanets and exoplanet candidates have been discovered and characterized due to the combined efforts of space-based and ground-based observational instruments and their surveys, such as by {\it Kepler} \citep{Borucki2010}, the Transiting Exoplanet Survey Satellite ({\it TESS}; \citealt{Ricker2015}), the Hungarian Automated Telescope ({\it HAT}; \citealt{Bakos2007}), the Hungarian Automated Telescope-South ({\it HATS};  \citealt{Bakos2013}), the Wide Angle Search for Planets ({\it WASP}; \citealt{Pollacco2006}), and the Kilodegree Extremely Little Telescope ({\it KELT}; \citealt{Pepper2007}), among others. While our understanding of exoplanet formation, system architectures, and evolution is far from complete, these surveys have revealed large populations of exoplanets that are very unlike our own solar system planets. In particular, many exoplanets with periods much shorter than that of Mercury have been discovered, such as Hot or Warm Jupiters, as well as previously unknown exoplanet types such as Super-Earths and Mini-Neptunes\footnote{See https://exoplanetarchive.ipac.caltech.edu/ for database of known exoplanets.}

Interestingly, short period exoplanets have been found around many stars much more massive than the Sun, such as A-type stars (e.g.,  \citealt{CollierCameron2010}, \citealt{Shporer2011}, \citealt{Szabo+2012}, \citealt{Gaudi+2017}), which are known to generally rotate rapidly for extended parts of their main sequence lifetime due to inefficient magnetic braking and tidal dissipation of their angular momentum \citep[e.g.,][]{Kraft1967,Ward+1976}. {Exoplanets transiting such stars can display interesting new phenomena that are less important for planets transiting less rapidly rotating stars, including distorted and asymmetric transits \citep{Barnes2009}, which can be used to measure the spin-orbit alignment of the planet \citep{Barnes2011}, gravity-darkened `seasons' \citep{Ahlers2020}, and orbital and/or spin precession \citep{Barnes+2013}}.  

{The nature of the precession depends on the magnitude of the angular momentum of the stellar spin $L_*$ compared to the magnitude of the angular momentum of the planetary orbit $L_p$ \citep[e.g.,][]{MD00}.   When $L_*\gg L_p$, the planet orbital angular momentum vector precesses about the stellar spin angular momentum vector. When $L_p\gg L_*$, the stellar spin angular momentum vector precesses about the planetary orbital angular momentum vector.  When $L_* \sim L_p$ then both the planetary orbital angular moment vector and the stellar spin angular moment vector mutually precess around the net angular momentum vector of the system.  Here we
focus on the first case, where the angular momentum of the stellar rotation dominates over the orbital angular momentum of its planet, which leads to nodal precession of the planet's orbit. This is generally the case for short period planets orbiting hot stars above the Kraft break (\citealt{Kraft1967}, \citealt{Ward+1976}, Figure \ref{fig:AM})}.

In some cases, precession is observable over relatively short observation time frames of only a few years \citep{johnson15,watanabe20,Borsa+2021,Stephan+2022}. The nodal precession of an exoplanet's orbit due to its host star's rapid rotation is a powerful tool to study the structural response of stars to the deformation caused by such rotation. So far, three rapidly precessing exoplanets have been observed and had their precession rates and stellar gravitational quadrupole moments measured \citep{Szabo+2012,johnson15,watanabe20,Borsa+2021,Stephan+2022}.

Beyond studying stellar structure, nodal precession also has an impact on our ability to detect exoplanets in the first place. Many surveys rely on the transit method, by which an exoplanet blocks out part of its host's light for some part of its orbit. The probability that a given planet will transit its host as seen from Earth is generally simply a function of the ratio of the host's radius divided by the orbital semi-major axis of the exoplanet, $R_*/a$. However, due to nodal precession, the relative orientation of exoplanet orbits for a fixed (e.g., Earth-based) observer can shift over time, allowing previously non-transiting exoplanets to transit at a later date, and vice-versa. As such, transit probability becomes not just a function of $R_*/a$, but also of the exoplanet's precession rate and overall time baseline that the system has been observed. Due to the continuing work of the various observational surveys, the observation time for significant parts of the sky has reached well over a decade.

In this work we investigate the time evolution of transit architectures due to nodal precession cause by rapidly rotating, oblate star, and the transit probability increase with increasing observation baselines. We provide equations that describe the time evolution of an exoplanet's impact parameter, $b$, and projected obliquity, $\lambda$, and outline which orbital architectures are most impacted by nodal precession, which ought to improve future studies of precessing exoplanets. Finally, we estimate how the nodal precession may impact the statistics of observed exoplanet system architectures, in particular regarding spin-orbit misalignment.

\section{Mathematical Methods}\label{sec:methods}

\begin{figure}[htbp]
    \centering
    \includegraphics[width=\linewidth]{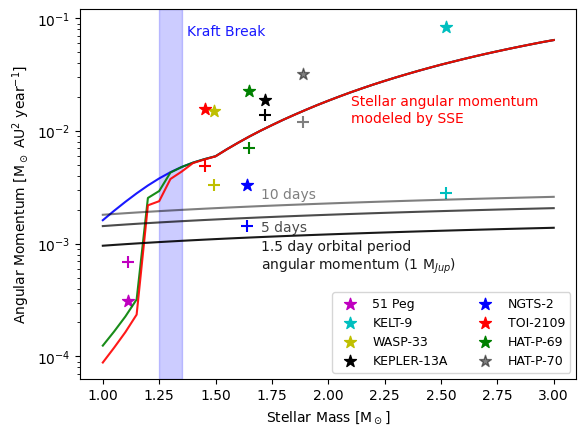}
    \caption{{\bf Angular momenta of rapidly rotating stars and their planets.} The figure shows the angular momenta of several rapidly rotating stars known to have short-period planets, compared to their planets' orbital angular momenta. The stellar rotational angular momenta, marked as stars, are calculated using a moment of inertia factor of $\alpha \sim 0.04$, a lower limit consistent with various models of stellar structure. The orbital angular momenta of the planets are marked by the plus signs. Model calculations using the stellar evolution code {\it SSE} \citep{Hurley+00} for the beginning, middle, and end of the main sequence evolution of stars covering the range of masses used in this figure are shown as blue, green, and red curves, respectively. The grey curves show three examples for the orbital angular momentum of a $1$~M$_{Jup}$ planet at a $1.5$, $5$, and $10$ day orbit. Note that Hot Jupiters orbiting stars above the Kraft break (to the right of the blue bar) generally have much lower orbital angular momenta than their host stars, as predicted by the model calculations. An example of a Hot Jupiter orbiting a star below the Kraft break, 51 Pegasi, for which the planet's angular momentum dominates, is given on the left side of the figure. As such, the nodal precession model described in this work appears generally valid for short-period planets orbiting stars above the Kraft break. Stellar and planetary parameters used to calculate the angular momenta are taken from observations \citep{MayorQueloz1995,Simpson+2010,Szabo+2012,vonEssen+2014,johnson15,Gaudi+2017,Raynard+2018,Zhou+2019,Wong+2021}.}
    \label{fig:AM}
\end{figure}

The impact parameter can be measured from the precise shape of the light curve alone (e.g., \citealt{Charbonneau:2000,Seager:2003,Carter:2008}), and is related to the inclination angle of the planet's orbital angular momentum vector against the line of sight, $i_p$, the stellar equatorial radius $R_*$, and orbital semi-major axis $a$ via the equation \begin{equation}\label{eq:b_i_p}
    b = (a/R_*) \cos i_p \ .
\end{equation} 
The quantity $a/R_*$ can be directly measured from the transit curve and radial velocity curve \citep{Seager:2003,Winn:2010,Carter:2008}, and thus it is possible to measure $i_p$ purely from observables.  
Disregarding the actual size of the transiting object, $b$ must be between the values of $-1$ and $1$ for transits to occur. 

\begin{figure}[htbp]
   \centering
    \includegraphics[width=\linewidth]{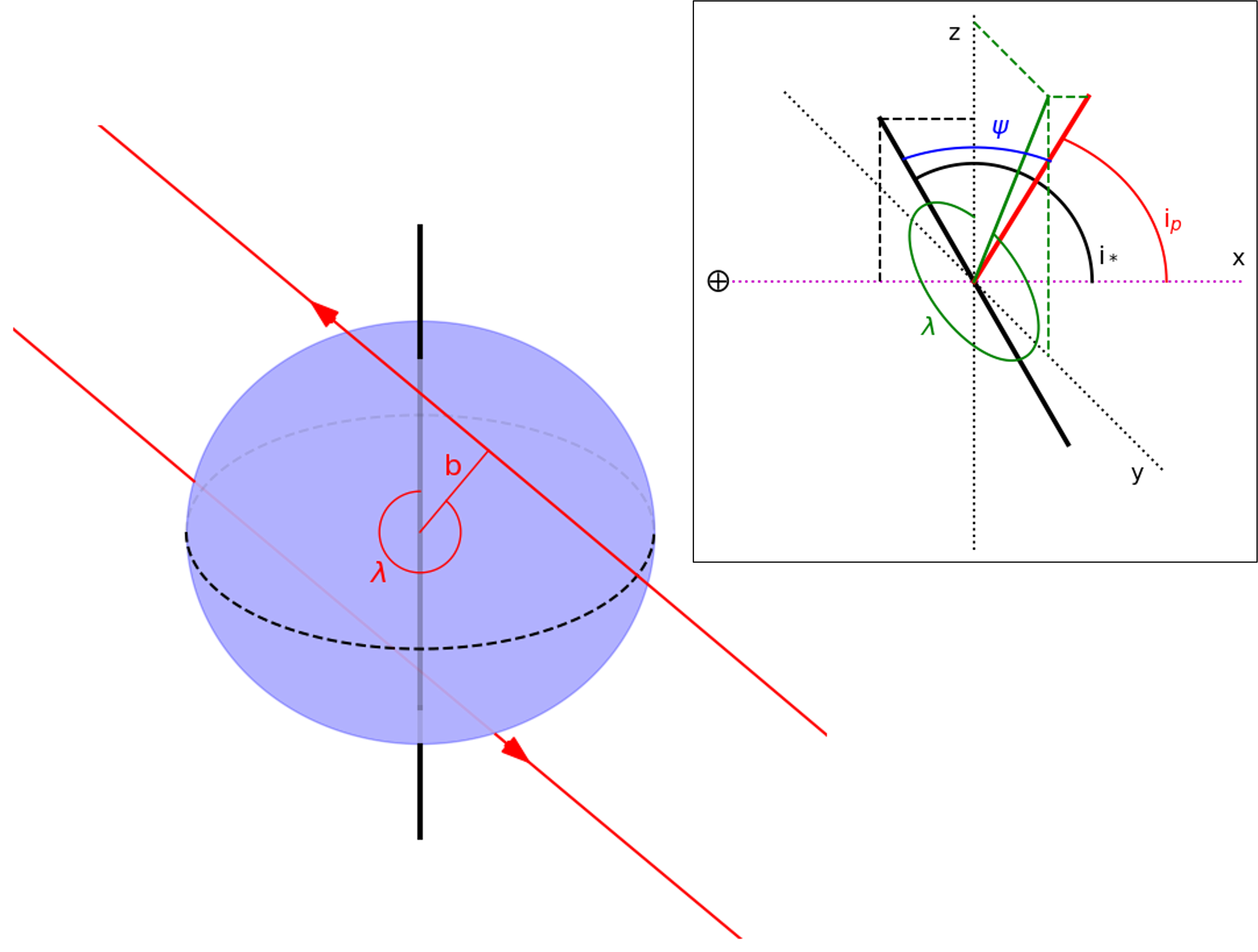}
    \caption{{\bf Observable transit geometry as defined in this study.} The figure presents an example orbital geometry of a rotating, oblate star being transited by a retrograde, inclined planet, shown from two perspectives. The left image shows the system from the observers perspective, with the star's spin axis (solid black line) oriented along the vertical axis. The star's spin axis is tilted towards the observer, as highlighted by its equator (dashed black line). The orbit's impact parameter $b$ and projected spin-orbit angle $\lambda$ are shown in red, as well as the orbital path (solid red line, direction marked with arrows). The right inset image shows an overview of the observable orbital angles, with the observer's line of sight being the x-axis (magenta dotted line, observer to the left, marked by $\bigoplus$). The tilt of the stellar spin axis (thick black line), i$_*$, is defined in relation to the line of sight. The red line shows the planet's orbital angular momentum vector, which forms an angle i$_p$ with the line of sight. The projection of the angular momentum vector onto the plane of the sky (the y-z plane), shown by the green line, forms the angle $\lambda$ with the z-axis, which is the projection of the stellar spin axis. The true spin-orbit angle $\psi$ is formed between the stellar spin axis and angular momentum vector and can be determined via Eq.~\ref{eq:psi_lambda}.
\label{fig:orientation}}
\end{figure}

The projected obliquity $\lambda$ is the two-dimensional projection of the true spin-orbit angle $\psi$. Their relation is described by the equation \begin{equation}\label{eq:psi_lambda}
    \cos \psi = \cos i_* \cos i_p + \sin i_* \sin i_p \cos \lambda \
\end{equation} \citep[e.g.,][]{iorio11}, where $i_*$ is the stellar spin angle versus the line of sight, defined such that $i_*=0\degree$ indicates that the star is viewed pole-on (see Fig.~\ref{fig:orientation} for an overview of the transit geometry). The projected obliquity can be measured a number of says, including the Rossiter-McLaughlin effect \citep{Rossiter:1924, McLaughlin:1924,Queloz:2000,Gaudi:2007}, Doppler Tomography \citep{CollierCameron2010, Johnson:2017}, gravity darkening \citep{Barnes2009,Barnes2011,Ahlers2020}, and starspots \citep{Desert:2011,Sanchis:2011b,Dai:2017}.  See \citet{Albrecht:2022} for a thorough review of measurements of the project obliquity of transiting exoplanets and references therein.  

These observable quantities are ultimately related to the orbital elements $\Omega$, the longitude of the ascending node, and $I$, the inclination of the orbital angular momentum versus the plane of the sky, via the equations \begin{equation}\label{eq:tanOmega}
    \tan \Omega = -\sin \lambda \tan i_p
\end{equation} and \begin{equation}\label{eq:cosI}
    \cos I = \cos \lambda \sin i_p \ ,
\end{equation} which allows for an alternate expression for $i_p$ in the form of \begin{equation}\label{eq:ip}
    \cos i_p = \sin I \cos \Omega \ .
\end{equation} In this work we are mostly interested in systems where the angular momentum of the stellar rotation dominates over the orbital angular momentum of its planet, which is generally the case for short period planets orbiting hot stars above the Kraft break \citep{Kraft1967, Ward+1976}, as we show in Fig.~\ref{fig:AM}. The long-term precession of $\Omega$ and $I$ of any planet with significantly smaller orbital angular momentum than its host star's rotational angular momentum can be described by the equations \citep[e.g.,][]{Iorio2016} \begin{equation}\label{eq:Omega_dot}
\begin{aligned}
    \dot{\Omega} = -\frac{3 \pi J_2 R_*^2}{2 P a^2} & \{ 2 \sin{i_*} \cos{i_*} \cos{2 I} \csc{I} \cos{\Omega} \\
    &- \cos{I}\left( 1 - 3 \sin^2{i_*} + \cos^2{i_*} \cos{2 \Omega} \right) \},
\end{aligned}
\end{equation}
and \begin{equation}\label{eq:I_dot}
\begin{aligned}
    \dot{I} = -\frac{3 \pi J_2 R_*^2}{P a^2} \cos{i_*}\sin{\Omega} \{& \sin{i_*}\cos{I} \\ &- \cos{i_*}\sin{I}\cos{\Omega} \},
\end{aligned}
\end{equation} with $P$ being the planet's orbital period and $J_2$ being the star's quadrupole gravitational moment. 

In the case that the rotating star is observed equator-on $(i_* = 90\degree)$, Equations \ref{eq:Omega_dot} and \ref{eq:I_dot} simplify to the straightforward expressions \begin{equation}\label{eq:Omega_dot_simple}
    \dot{\Omega}_{eq} = -\frac{3 \pi J_2 R_*^2}{P a^2}  \cos{I_{eq}} ,
\end{equation} and \begin{equation}\label{eq:I_dot_simple}
    \dot{I}_{eq} = 0 ,
\end{equation} which are the standard equations for nodal precession in the frame of the stellar spin, for which $I=I_{eq}=\psi$.

\begin{figure}[htbp]
    \centering
    \includegraphics[width=\linewidth]{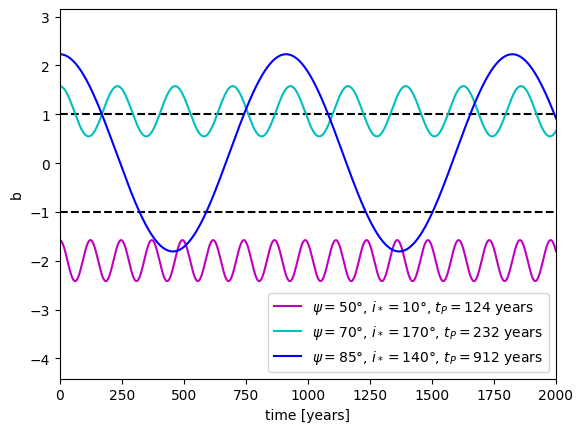}
    \caption{{\bf Evolution of the impact parameter $b$ over time.} The figure shows the time evolution of $b$ following Equation \ref{eq:b_time} for three example geometries with values of $i_*$ and $\psi$ as shown in the figure legend. The three examples share the same overall system parameters, $a/R_* = 3.153$, $J_2 = 3.38\times10^{-4}$, and $P = 127969~{\rm s}\simeq 1.48~{\rm days}$, representative of the KELT-9 system \citep{Stephan+2022}. The blue curve was produced using parameters similar to the best-fit solution for the precession of KELT-9b from \citet{Stephan+2022} and is an example of a large amplitude oscillation of $b$ with clear transit signals for extended times. The cyan curve shows an example where the planet would oscillate closely around a grazing transit at $b=1$, while the magenta curve shows an example of a planet that would never transit. A numerical integration of Equations \ref{eq:Omega_dot} and \ref{eq:I_dot} yields identical curves to within a phase offset.} 
    \label{fig:transit_examples}
\end{figure}

While Equations \ref{eq:Omega_dot} and \ref{eq:I_dot} are not easily integratable over time, Equations \ref{eq:Omega_dot_simple} and \ref{eq:I_dot_simple} have straightforward integral solutions in \begin{equation}\label{eq:Omega_simple}
    {\Omega}_{eq}(t) = \Omega_{eq,0} -\left(\frac{3 \pi J_2 R_*^2}{P a^2}  \cos{I_{eq,0}}\right) t 
\end{equation} and \begin{equation}\label{eq:I_simple}
    {I}_{eq}(t) = I_{eq,0} = \psi = {\rm const} .
\end{equation} From this it becomes clear that the precession of $\Omega_{eq}$ has a circulation period of \begin{equation}\label{eq:prec_period}
    t_P = \frac{2 P a^2 }{3 J_2 R_*^2 |\cos{\psi}|} ,
\end{equation} applicable to $\Omega$ and $I$ in all reference frames. Furthermore, we can construct a new expression for the time evolution of impact parameter $b$ by combining Equations \ref{eq:ip}, \ref{eq:Omega_simple}, and \ref{eq:I_simple}. As a first step, consider that the maximum and minimum values of $b$ are reached when $\cos{\Omega}=\pm 1$ and that $I$ can only oscillate between values of $\psi - (90\degree - i_*)$ and $\psi + (90\degree - i_*)$. Since $\Omega$ and $I$ oscillate with the same period, the extreme values of $b$ are thus given by \begin{equation}\label{eq:b_step0}
    b_{extrema} = \pm\frac{a}{R_*} {\sin{(\psi\mp(90\degree -i_*))}},
\end{equation} which can be simplified to \begin{equation}\label{eq:b_step1}
    b_{extrema} = -\frac{a}{R_*} {\cos{(\psi \pm i_*)}}.
\end{equation} The value of $b$ will oscillate between these extreme values with a period of $t_P$, as this is the oscillation period in all reference frames. We thus construct the equation \begin{equation}\label{eq:b_step2}
\begin{aligned}
    b(t) = \frac{a}{R_*} \biggl\{&\frac{-\cos{(\psi+i_*)}-\cos{(\psi-i_*)}}{2} \\ +&\frac{-\cos{(\psi+i_*)}+\cos{(\psi-i_*)}}{2} \\& \times\cos{\left(\frac{2\pi}{t_P}\times t\right)}\biggr\}.
    \end{aligned}
\end{equation} Finally, via the trigonometric identities concerning the sum of angles, we arrive at the expression \begin{equation}\label{eq:b_time}
\begin{aligned}
    b(t) = \frac{a}{R_*} \biggl\{ &-\cos{\psi}\cos{i_*} \\& +\sin{\psi}\sin{i_*}\cos{\left(\frac{2\pi}{t_P}\times t\right)}\biggr\}.
    \end{aligned}
\end{equation} {By numerically integrating Equations \ref{eq:Omega_dot} and \ref{eq:I_dot}, we have verified that this expression describes the evolution of $b$ for any system and any observing geometry, assuming one can derive the true spin-orbit angle, $\psi$. However, the phase of the oscillation has to be adjusted by the correct choice of $t$ depending on observations.} The time derivative of $b$ is, consequently,\begin{equation}\label{eq:bdot_time}
    \dot{b}(t) = -\frac{a}{R_*}\frac{2\pi}{t_P} \sin{\psi}\sin{i_*}\sin{\left(\frac{2\pi}{t_P}\times t\right)}.
\end{equation} For any relevant host star with persistent physical characteristics, the time evolution of $b$ and $\dot{b}$ thus purely depends on the angle of the stellar spin axis versus the line of sight, $i_*$, and the true spin-orbit angle, $\psi$. Figure \ref{fig:transit_examples} shows several examples of the evolution of $b$ following Equation \ref{eq:b_time}. For $\lambda$ a similar time evolution expression can be constructed. First, by rearranging Equation \ref{eq:psi_lambda}, we obtain\begin{equation}\label{eq:lambda_1}
    \cos{\lambda} = \frac{ \cos{\psi} - \cos{i_p} \cos{i_*} }{\sin{i_*}\sin{i_p}} .
\end{equation} Since $\cos{i_p}=(R_*/a)\times b$, we thus arrive at the equation\begin{equation}\label{eq:lambda_time}
    \cos{\lambda(t)} = \frac{ \cos{\psi} - \frac{R_*}{a} b(t) \cos{i_*} }{\sin{i_*}\sqrt{1-\left[\frac{R_*}{a} b(t)\right]^2}} .
\end{equation} Here, we primarily focus on the time evolution of $b$ as described by Equation \ref{eq:b_time}.

\subsection{Transit Probability Increase due to Nodal Precession}\label{subsec:prec}

\begin{figure}[htbp]
    \centering
    \includegraphics[width=\linewidth]{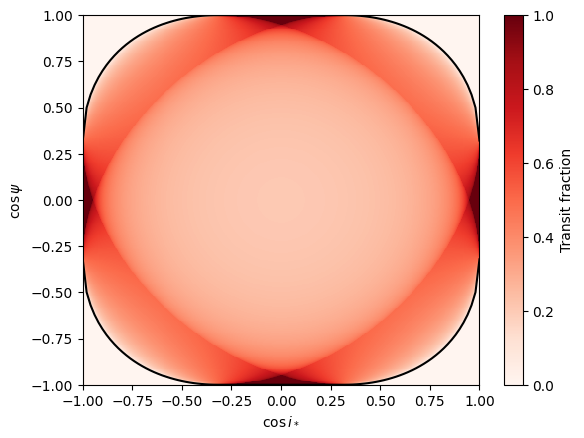}
    \caption{{\bf Fractions of the precession periods that planets will be transiting their host stars as a function of $i_*$ and $\psi$.} The figure shows the likelihood for a planet to transit its host star in the $\cos{i_*}$-$\cos{\psi}$ parameter space, as determined by Equation \ref{eq:b_time} as a fraction of its precession period. Only the parameter space inside the black outline allows transits to occur at some point of the system's lifetime (see Equations \ref{eq:b_max}, \ref{eq:b_min}). The figure serves to highlight the general shape of this function, the exact numerical values are dependent on the stellar and orbital parameters (see Equations \ref{eq:prec_period} and \ref{eq:b_time}). In this example, $a/R_* = 3.153$, $J_2 = 3.38\times10^{-4}$, and $P = 127969~{\rm s}\simeq 1.48~{\rm days}$, representative of KELT-9b \citep{Stephan+2022}. The average transit fraction is $0.317=R_*/a$.} 
    \label{fig:transit_fraction}
\end{figure}

The expression for $b$ as shown by Equation \ref{eq:b_time} enables us to recognize certain geometric relations between $i_*$, $\psi$, and the possibility of transits that make intuitive sense. If the star is viewed pole-on $(i_* = 0\degree)$, the orbit's impact parameter will never change, even though the planet is precessing. As such, only certain angles of $\psi$ will result in transits, which consequently will then always be observable. Alternatively, if the star is viewed equator-on $(i_* = 90\degree)$, any orientation of $\psi$ will result in transits for at least some fraction of the precession period, though precession speed will, of course, still depend on $t_P\propto1/\cos{\psi}$. As such, if the planet's orbital plane is aligned with the star's equator $(\psi = 0\degree)$, while the precession period, $t_P$, will be at its shortest, precession will not be observable, as the amplitude of the change in $b$ will be zero. If the planet's orbital plane is exactly aligned with the stellar rotation axis $(\psi = 90\degree)$, the precession period approaches infinity. Figure \ref{fig:transit_fraction} shows how these geometric relations translate into transit fractions over the course of the precession periods. These fractions can be calculated by determining the times when the impact parameter $b$ crosses the values of $-1$ or $1$, via the expression \begin{equation}\label{eq:time_b}
    t(b) = \frac{t_P}{2\pi} \arccos{\left( \frac{ b\frac{R_*}{a} + \cos{\psi} \cos{i_*}}{\sin{\psi} \sin{i_*} } \right)},
\end{equation} and comparing it to the precession timescale. The boundary between the never-transit and sometimes-transit regions in Figure \ref{fig:transit_fraction} is formed by geometries where the minimum value of $b$ during precession is exactly equal $1$ or the maximum value of $b$ is equal $-1$, such that\begin{equation}\label{eq:b_max}
    b_{max} = \max\left(-\frac{a}{R_*}\cos{(\psi\pm i_*)}\right) = -1,
\end{equation} and \begin{equation}\label{eq:b_min} 
    b_{min} = \min\left(-\frac{a}{R_*}\cos{(\psi\pm i_*})\right) = 1.
\end{equation}

\begin{figure}[htbp]
    \centering
    \includegraphics[width=\linewidth]{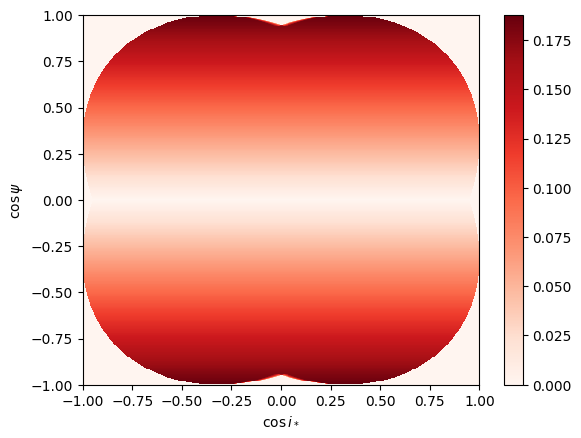}
    \caption{{\bf Increase of the transit likelihood due to extended observation time baseline.} The figure shows how an extended observation period increases the transit likelihood by allowing significant nodal precession to occur. As such, the increase is primarily a function of $\cos{\psi}$, as it depends on the precession period, $t_P$. The system parameters are identical to the one in Fig.~\ref{fig:transit_fraction}. The observation baseline used is $15$ years, motivated by the {\it Kepler} and {\it TESS} missions. On average, the observation likelihood for a system in this example increases by about $8\%$, while for the fastest precessing systems ($\psi$ close to $0\degree$ or $180\degree$), the transit likelihood increases by nearly $19\%$.} 
    \label{fig:transit_increase}
\end{figure}

While the transit fractions over precession period as shown in Fig.~\ref{fig:transit_fraction} give us a sense for what values of $\psi$ and $i_*$ are more likely to be observed for any particular system, they do not change the overall transit likelihood for a random collection of systems over a short observation timescale. This ``instantaneous'' transit likelihood still follows $P_{transit} = R_*/a$. However, due to nodal precession, the transit likelihood does indeed change given a longer observation time period. The {\it Kepler} and {\it TESS} missions give us such an extended observation period, for a portion of the sky, on the order of $15$~years, assuming TESS continues at least until $2024$~CE. As such, the transit likelihood for any particular system changes to \begin{equation}
    P_{transit, new} = \frac{R_*}{a} + \frac{t_{obs}}{t_P},
\end{equation} with $t_{obs}$ being the observation time period. $P_{transit, new}$ naturally reaches a maximum of $1$ when $t_{obs}/t_P$ reaches $1-R_*/a$. The transit likelihood increase is thus also a function of $\psi$, with values of $\psi$ close to $0\degree$ or $180\degree$ resulting in the shortest precession periods, experiencing the largest increase of their transit likelihood. Fig.~\ref{fig:transit_increase} shows this increase in transit likelihood in the same phase space as used in Fig.~\ref{fig:transit_fraction}.

We note here that the increase of the observation likelihood is small for most exoplanets. Only very short-period planets orbiting hot, rapidly rotating stars (which thus should posses a significant equatorial bulge and large gravitational quadrupole moment) will experience nodal precession fast enough to significantly affect the likelihood. For example, for an exoplanet like WASP-33b \citep{johnson15,watanabe20,Stephan+2022}, assuming an isotropic likelihood distribution for $\cos{\psi}$ and $\cos{i_*}$, a $15$ years observation time period would increase the overall observation likelihood by $1~\%$, and by about $3~\%$ for the most favorable possible orientations. For an exoplanet like KELT-9b \citep{Stephan+2022}, however, the increase is substantially higher over the same time frame, with an $8~\%$ average increase and a $19~\%$ increase for the most favorable orientations (see Fig.~\ref{fig:transit_increase}).

\section{Discussion}\label{sec:disc}

The increase in transit likelihood, as described in the previous section, is dependent on the observation time frame, spin-orbit angle $\psi$, stellar $J_2$ value, and orbital period of the planet, and creates a set of biases for transit observations. In general terms, exoplanets orbiting with a spin-orbit angle $\psi$ close to $0\degree$ or $180\degree$, around stars with large $J_2$ values, on short orbits, are more likely to be observed as transiting planets at some point during any given observation time baseline, with the relative proportion of such planets increasing with increasing time baseline. While these effects ought to be small to negligible for most exoplanets, for certain classes the effects may be significant, which we outline here.

\subsection{Hot Jupiters orbiting Hot Stars}\label{subsec:HJs}

As outlined in Section \ref{sec:methods}, hot stars with short spin periods and large gravitational quadrupole moments are the primary environment to cause rapid nodal precession of close-in planetary orbits. So far, three Hot Jupiters orbiting such stars have measured precession periods, namely Kepler-13Ab  \citep[about $500$~years, e.g.,][]{Szabo+2012}, WASP-33b \citep[in the range of $800$ to $1500$~years, e.g.,][]{johnson15,watanabe20,Borsa+2021,Stephan+2022}, and KELT-9b \citep[about $900$~years,][]{Stephan+2022}. These planets orbit their host stars with a range of obliquity values and the stars have $J_2$ values in the approximate range of $6\times10^{-5}$ to $4\times10^{-4}$. As such, the transit likelihood increases for these three known precessing Hot Jupiters over the approximately $15$~year long observation time frame of the {\it Kepler} and {\it TESS} missions are non-negligible. 

For a planet like Kepler-13Ab, the base transit probability, derived from its $a/R_*$ value of $\sim4.44$ \citep[e.g.,][]{Shporer+2014}, is about $22.5\%$. Given its rapid precession, over a $15$~years observation time frame this probability increases by $3\%$ to about $25.5\%$. For KELT-9b, the base transit probability of $31.7\%$ \citep[e.g.,][]{Gaudi+2017} increases by $1.7\%$ to $33.4\%$, and for WASP-33b the base transit probability of $26.4\%$ \citep[e.g.,][]{CollierCameron2010} is increased by between $1$ and $1.9\%$ up to between $27.4$ and $28.3\%$. 

While these increases are overall comparatively small and do not change the transit likelihoods of these planets in a qualitatively significant way, they highlight that there potentially exists a systemic underestimation of the transit likelihoods of Hot Jupiters that may affect population-level studies, at least around massive, hot, rapidly rotating stars. In particular, applying the equations outlined in Section \ref{sec:methods}, one can estimate the potential transition likelihood increases for the most rapidly precessing orientations of $\psi$. For the example of a planet like KELT-9b, the potential transit likelihood increase would be up to about $19\%$, with an average increase for all possible orientations in the $\cos{\psi}$, $\cos{i_*}$ phase space of about $8\%$ (see Figure \ref{fig:transit_increase}). Such an increase would be on the order of nearly $1/3$ to $2/3$ of the base transit probability for a planet like KELT-9b, potentially even on the order of $1$ or larger for a planet like Kepler-13Ab, significantly impacting observational statistics, assuming that the distributions of $\cos{\psi}$ and $\cos{i_*}$ are truly isotropic. In fact, as more rapidly precessing planets tend to have orbits more aligned with their host star's equator, this effect creates an observational bias against misaligned, slowly precessing planets. {\bf As such, statistical estimates of the distribution of aligned versus misaligned Hot Jupiter orbits will tend to overestimate the inherent fraction of aligned systems}. This effect will become increasingly important as the time baseline over which a large fraction of the sky has been observed increases with various future missions and ground-based follow-up surveys.

\subsection{Comparison to Circumbinary Planets}\label{subsec:CBPs}

Much of the basic precession physics presented in this work is similar in nature to the precession observed for circumbinary planets (CBPs) \citep[e.g.,][]{Martin2017}. However, the timescales for the precession of close-in orbiting CBPs is generally significantly shorter than that of short-period planets around hot stars and is on the order of decades rather than centuries or millennia, in part due to the much more significant angular momentum of a tight binary star that is driving the planetary orbit's nodal precession. However, the transit geometry is more complicated given by the binary nature of the to-be-transited host, making analysis of the transit evolution more complex. Furthermore, the speed of the precession is significantly impacted by the host stars' internal structure or tidal mechanics, but mostly by the mass and orbital configuration of the binary. As such, the two cases complement each other for these separate planetary populations.

\section{Summary and Conclusions}\label{sec:summary}

In this work we have provided a description of the time evolution of exoplanet transits due to nodal precession caused by rapidly rotating host stars, generally valid for short-period planets orbiting stars above the Kraft break. We derived analytical expressions for the time evolution of the impact parameter $b$ (see Equations \ref{eq:b_time} and \ref{eq:bdot_time}), and estimated the impact of nodal precession and orbital architectures on increasing transit likelihoods. We can draw two major conclusions based on our investigation:\begin{enumerate}
  \item For the most rapidly precessing exoplanets, such as Hot Jupiters orbiting rapidly rotating massive stars, the observation time frame covered by {\it Kepler} and {\it TESS} for parts of the sky has increased the transit likelihoods by a non-negligible amount on the order of a few to $\sim10~\%$. As such, studies that estimate planet occurrence rates based on these surveys should take precession into account for their calculations.
  \item Due to the dependence of precession rates on the orbital orientation of the exoplanets, in particular the spin-orbit angle $\psi$ as shown in Equation \ref{eq:prec_period}, planets that are more aligned with their host stars' spins will be more likely to be observed over time than misaligned planets. This effect is especially important for the study of Hot Jupiters, as tidal migration models of their formation generally predict misaligned or randomly aligned orbits. Furthermore, different alignment distributions have been observed for low mass vs. high mass stars, pointing to either different formation mechanisms or different tidal re-alignment efficiencies between these groups of stars, as generally expected due to the Kraft break \citep[e.g.,][]{Kraft1967,Ward+1976}. As such, the dependence of transit likelihood on orbital orientation may skew conclusions drawn based on observations alone.
\end{enumerate} Additionally, as significant nodal precession ought to be limited to short-period planets orbiting hot, young, rapidly-rotating, massive stars, observations may also eventually overestimate planet occurrence rates for these types of stars compared to lower-mass and older, slowly rotating ones.

In conclusion, the impact of nodal precession on transit likelihoods will have to be accounted for when attempting to derive accurate short-period exoplanet occurrence rates and the distribution of spin-orbit alignment angles. While the effect is currently relatively small, on the order of a few percent on average, for certain architectures the effect should already be on the order of $10$ to $20\%$, thus potentially skewing derived distribution functions.

\section*{Acknowledgments}
We would like to thank the anonymous referee for helpful suggestions. 
We thank Marshall C. Johnson and David V. Martin for helpful comments and discussions. A.P.S. acknowledges partial support from the President's Postdoctoral Scholarship from the Ohio State University and the Ohio Eminent Scholar Endowment. A.P.S. and B.S.G. acknowledge partial support by the Thomas Jefferson Chair Endowment for Discovery and Space Exploration.  





\bibliographystyle{aasjournal}
\bibliography{Kozai2}{} 

\end{CJK*}

\end{document}